\documentclass[twocolumn,prl,aps,showpacs]{revtex4}         
\usepackage{epsfig}
\catcode`\ä = \active \catcode`\ö = \active \catcode`\ü = \active
\catcode`\Ä = \active \catcode`\Ö = \active \catcode`\Ü = \active
\catcode`\ß = \active
\defä{\"a}
\defö{\"o}
\defü{\"u}
\defÄ{\"A}
\defÖ{\"O}
\defÜ{\"U}
\defß{\ss}

\begin{document}

\title{Sodium Bose-Einstein Condensates in the F=2 State in a
Large-volume Optical Trap}
\author{A. G\"orlitz\cite{byline}, T.~L. Gustavson\cite{byline2}, A.~E. Leanhardt, R.~L\"ow\cite{byline}, A.~P. Chikkatur, S. Gupta,
S. Inouye\cite{byline3}, D.~E. Pritchard and W. Ketterle}
\affiliation{Department of Physics, MIT-Harvard Center for
Ultracold Atoms, and Research Laboratory of Electronics, \\
Massachusetts Institute of Technology, Cambridge, MA 02139}
\date{\today}

\begin{abstract}
We have investigated the properties of Bose-Einstein condensates
of sodium atoms in the upper hyperfine ground state in a purely
optical trap. Condensates in the high-field seeking
$|$F=2,\,m$_F$=-2$\rangle$ state were created from initially
prepared $|$F=1,m$_F$=-1$\rangle$ condensates using a one-photon
microwave transition at 1.77\,GHz. The condensates were stored in
a large-volume optical trap created by a single laser beam with an
elliptical focus. We found condensates in the stretched state
$|$F=2,\,m$_F$=-2$\rangle$ to be stable for several seconds at
densities in the range of $10^{14}$\,atoms/cm$^{3}$. In addition,
we studied the clock transition $|$F$=1$,
m$_F$=0$\rangle$\,$\rightarrow$\,$|$F=2, m$_F$=0$\rangle$ in a
sodium Bose-Einstein condensate and determined a density-dependent
frequency shift of $(2.44\pm 0.25) \times 10^{-12}$
\,Hz\,cm$^{3}$.
\end{abstract}
 \vskip1pc \maketitle

So far, Bose-Einstein condensation in dilute atomic gases
\cite{ande95,davi95bec,brad97bec,corn00,modu01} has been achieved
in all stable bosonic alkali isotopes except $^{39}$K and
$^{133}$Cs, as well as in atomic hydrogen \cite{frie98} and
metastable helium \cite{robe01,pere01}. The physics that can be
explored with Bose-Einstein condensates (BEC) is to a large extent
governed by the details of interatomic interactions. At ultra-low
temperatures, these interactions  not only vary significantly from
one atomic species to another but can also change significantly
for different internal states of a single species. While in
$^{87}$Rb, only minor differences of the collisional properties
are observed within the ground state manifolds, in $^{7}$Li, the
magnitude of the scattering length differs by a factor of five
between the upper and the lower hyperfine manifold and even the
sign is inverted \cite{schr01}. The behavior of $^{23}$Na with a
scattering length of 2.80\,nm in the $|$F=1,m$_F$=$\pm$1$\rangle$
states and 3.31\,nm in the  $|$F=2,m$_F$=$\pm$2$\rangle$ states
\cite{samu00} is intermediate between these two extreme cases.
Thus, sodium might provide a system in which the study of BEC
mixtures of states with significantly differing scattering length
is possible. Such a mixture would be a natural extension of
earlier work on spinor condensates in $^{87}$Rb
\cite{myat97,hall98dyn} and in the F=1 manifold of $^{23}$Na
\cite{sten98spin,mies99meta}.

In this Letter, we report the realization of Bose-Einstein
condensates of $^{23}$Na in the upper F=2 hyperfine manifold in a
large-volume optical trap \cite{f2mag}. In $^{87}$Rb, condensates
in both the F=1 and F=2 states had been achieved by loading atoms
in either state into a magnetic trap and subsequent evaporative
cooling. In contrast, sodium BECs have previously only been
produced in the F=1 state. Early attempts at MIT and NIST to
evaporatively cool sodium in the F=2 state were discontinued since
the evaporative cooling scheme proved to be more robust for the
F=1 state. Instead of developing an optimized evaporation strategy
for F=2 atoms in a magnetic trap, we took advantage of an optical
trap which traps atoms in arbitrary spin states \cite{stam98odt}.
After producing F=1 condensates and loading them into an optical
trap, we transferred the population into the F=2 manifold using a
single-photon microwave transition at 1.77\,GHz. We found that a
BEC in the stretched
 $|$F=2,\,m$_F$=-2$\rangle$ state is stable on timescales of seconds at densities
 of a few
 $10^{14}\, \text{atoms/cm}^3$. Simultaneous trapping of condensates in the
 $|$2,-2$\rangle$ and $|$1,-1$\rangle$ states for several seconds was also achieved.
 In contrast, at the same density, a condensate in
 the $|$2,0$\rangle$ state decays within milliseconds. Nevertheless,
 we were able to observe the so-called clock transition
  $|1,0\rangle \, \rightarrow \, |2,0\rangle$ in a BEC, which is to lowest order
  insensitive to stray magnetic fields.
  By taking spectra of this transition at various condensate densities, we were able to measure
 a density-dependent frequency shift of $(2.44\pm 0.25) \times 10^{-12}$\,Hz cm$^{3}$.

The basic setup of our experiment is described in
\cite{kett99var,gorl011D2D} and is briefly summarized here. We
have prepared condensates of more than $4 \times 10^7$ $^{23}$Na
atoms in a so-called `clover-leaf' magnetic trap with trapping
frequencies of $\nu_x=16\,$Hz and $\nu_y=\nu_z = 160\,\text{Hz}$
by radiofrequency evaporation for 20\,s. After preparation of the
condensate in the $|1,-1\rangle$ state, the radial trapping
frequencies were adiabatically lowered by a factor of 5 to
decompress the condensate. Subsequently, an optical trapping
potential was superimposed on the condensate by slowly ramping up
the light intensity. After turning off the remaining magnetic
fields, nearly all atoms were loaded into the large-volume optical
dipole trap. The resulting peak density reached $5 \times
10^{14}$\,atoms/cm$^3$, slightly higher than the density in the
magnetic trap.

The large-volume optical trap was realized by shaping the output
of a Nd:YAG laser (typically 500 mW at 1064 nm) with cylindrical
lenses leading to an elliptical  focus with an aspect ratio of
approximately 25. At the location of the condensate, the focal
size was $\approx 20 \mu$m along the tight axis resulting in an
optical trapping potential with typical trap frequencies of
$\nu_x$\,=\,13\,Hz axially and $\nu_y$\,=\,36\,Hz and
$\nu_z$\,=\,850\,Hz transversely. The trap axis with the largest
trapping frequency was oriented vertically to counteract gravity.
The pancake shape of the trap, which we had recently used to
create (quasi-) 2D condensates \cite{gorl011D2D}, provided a much
larger trapping volume than our previous cigar-shaped optical
traps \cite{stam98odt,gust02} and thus significantly larger
condensates could be stored.

\begin{figure}
\epsfig{file=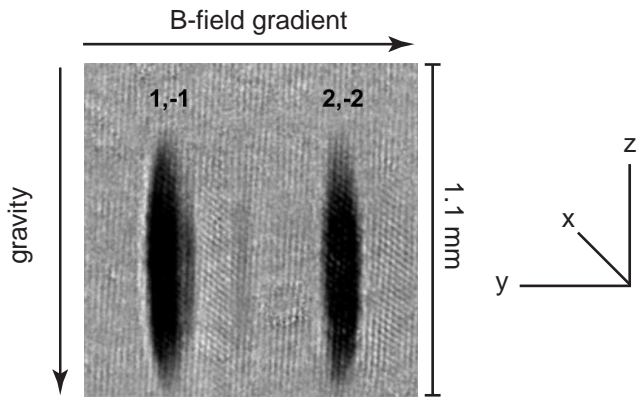}  \caption{\label{fig:exp_scheme}Sodium
condensates in the $|$1,-1$\rangle$ and $|$2,-2$\rangle$ state,
30\,ms after release from the trap. After preparation of the
mixture the atoms were held in the optical trap for 1\,s. The
horizontal separation of the spin states is due to application of
a magnetic-field gradient during expansion.}
\end{figure}

Optically trapped condensates were observed by absorption imaging
 on the closed
$|$F=2,m$_F$=-2$\rangle\,\rightarrow\,|$F$'$=3,m$'_F$=-3$\rangle$
 cycling transition at 589\,nm after sudden release from the trap,
 using light propagating parallel to the trap laser. The
ballistic expansion time was typically 30\,ms, after which the
vertical size of the condensate had increased by more than a
factor of 100 while the horizontal expansion was less than a
factor of two. To make sure that atoms in both the F=1 and the F=2
manifold could be detected simultaneously, a short laser pulse
resonant with the F=1\,$\rightarrow$\,F$'$=2 transition was
applied to pump all atoms into the F=2 manifold. State-selective
detection could be achieved by applying a magnetic field gradient
of several G/cm during the free expansion of the atomic cloud,
leading to a spatial separation of spin states which differ in the
orientation of the magnetic moment (Fig.\ \ref{fig:exp_scheme}).

\begin{figure}
\epsfig{file=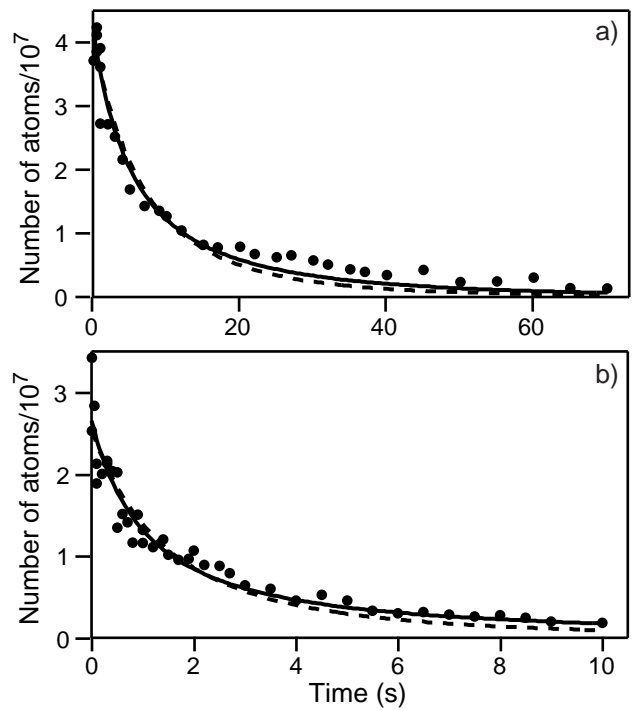} \caption{\label{fig:lifetime}Lifetime
measurement of sodium BECs in the $|$1,-1$\rangle$ (a) and
$|$2,-2$\rangle$ (b) states in a large-volume optical trap with an
initial peak density of $\approx 5 \times 10^{14}$\,atoms/cm$^3$.
The number of atoms is determined from the measured size of the
clouds after ballistic expansion. The lines are fits where either
only three-body loss (solid) or two-body loss (dashed) has been
assumed in addition to a one-body loss rate of 0.029\,s$^{-1}$.}
\end{figure}

In order to test the intrinsic stability of the optical trap, we
first investigated the lifetime of condensates in the
$|$1,-1$\rangle$ state as shown in Fig.\ \ref{fig:lifetime}\,a).
Even after 70\,s of dwell time, more than 10$^6$\,atoms remained
in the condensate. Generally, the decay of the number of atoms $N$
in the condensate can be modelled by the rate equation
\begin{equation}
\frac{dN}{dt}\,=\,-k_1 N \,-\, k_2 N \langle n \rangle\, - \, k_3
N\langle n^2 \rangle\ \, , \label{Eq:N_decay}
\end{equation}
\noindent where $k_1,k_2,k_3$ are the one-, two- and three-body
loss coefficients and $n$ is the condensate density. By setting
either $k_2$ or $k_3$ to zero in Eq.\ \ref{Eq:N_decay}, analytical
fitting functions for the decay of the condensate number can be
derived. A fit to the data in Fig.\ \ref{fig:lifetime}\,a) with
$k_2 = 0$ and a one-body loss rate $k_1 = 0.029$ s$^{-1}$
(determined by an exponential fit to all data points with $t \geq
40$\,s) yields a three-body loss coefficient of $k_3 = (1.57\pm
0.17 \pm 0.55) \times 10^{-30}$\,cm$^6$\,s$^{-1}$, in agreement
with the previously published value \cite{stam98odt}. Here, the
first error is the statistical error of the fit and the second
error represents systematic uncertainties originating from the
measurements of the cloud sizes and the determination of $k_1$.
Since the densities in the present work are lower than in previous
work, two-body loss processes might also contribute. A fit where
$k_3$ was set to zero yielded a two-body rate coefficient of $k_2
= (3.98\pm 0.47 \pm 0.81) \times 10^{-16}$\,cm$^3$\,s$^{-1}$,
larger than theoretical predictions \cite{boes96dipolar}. The fits
indicate that the observed loss is predominantly due to three-body
collisions, though two-body processes cannot be ruled out.

Condensates in the $|$2,-2$\rangle$ state were produced by
applying a microwave pulse at 1.77\,GHz to an optically trapped
$|$1,-1$\rangle$ condensate. By varying power and duration of the
microwave pulse, we were able to adjust the ratio of atoms
transferred into the $|$2,-2$\rangle$ state between 0 and 100\,\%.
Fig.\ \ref{fig:lifetime}\,b) shows a measurement of the lifetime
after complete transfer into the $|$2,-2$\rangle$ state. The
lifetime in the $|$2,-2$\rangle$ state is still on the order of
seconds but it is significantly shorter than the lifetime of a
$|$1,-1$\rangle$ condensate (Fig.\ \ref{fig:lifetime}\,a). The
reduced lifetime can be attributed to much larger three- and/or
two-body loss rates. Using the solutions of Eq.\ \ref{Eq:N_decay}
we deduce rate coefficients for the atom loss, assuming that only
one process is responsible for the loss. Thus, we obtain as upper
bounds $k_2\,=\,(2.93 \pm 0.28 \pm 0.29 ) \times
10^{-15}$\,cm$^3$\,s$^{-1}$ and $k_3\,=\,(1.53\pm 0.13 \pm 0.32)
\times 10^{-29}$\,cm$^6$\,s$^{-1}$. Both values are in reasonable
agreement with theoretical predictions
\cite{moer96coll,moer96reco}. Though, at typical densities, the
decay rate in the F=2 state is roughly an order of magnitude
larger than in the F=1 state, it should still be compatible with
direct condensation in the F=2 manifold, provided that the loss
coefficients for the magnetically trapable $|$2,+2$\rangle$ state
are similar to those for the $|$2,-2$\rangle$ state.

By transferring only part of the atoms into the upper hyperfine
manifold we could also observe mixtures of condensates in the
$|$1,-1$\rangle$ and $|$2,-2$\rangle$ states (see Fig.\
\ref{fig:exp_scheme}). In the presence of small magnetic field
gradients, we observed a rapid spatial separation of the two
components in a time shorter than 100\,ms due to the fact that the
$|$1,-1$\rangle$ state is low-field seeking while the
$|$2,-2$\rangle$ state is high-field seeking. During the
separation, strong density modulations in both components were
observed, which could be attributed to tunnelling processes
playing a role in the separation process \cite{stam99tun}.
Afterwards, the two components lived almost independently side by
side in the trap and the individual lifetimes were not
significantly affected. When we tried to compensate all stray
magnetic field gradients, we still found that in steady state the
two components tend to separate, i.e. we observed domains with
only one component \cite{mies99meta}. This indicates that the two
states are intrinsically not miscible. While we found $^{23}$Na
BECs in the $|$2,-2$\rangle$ state as well as mixtures of
$|$1,-1$\rangle$ and $|$2,-2$\rangle$ condensates to be stable for
several seconds, non-stretched states in the F\,=\,2 manifold as
well as F\,=\,1, F\,=\,2 mixtures with $|$m$_1$ + m$_2| \neq 3$
decayed within several ms for typical condensate densities on the
order of 10$^{14}$\,atoms/cm$^3$. This fast decay is probably due
to (two-body) spin-relaxation which is strongly suppressed in
$^{87}$Rb but occurs with rate constants on the order of
10$^{-11}$\,cm$^3$s$^{-1}$ in $^{23}$Na \cite{moer96coll}.

A particularly interesting transition within the electronic ground
state of alkali atoms is the magnetic-field insensitive transition
$|$F,0$\rangle$\,$\rightarrow\,|$F+1,0$\rangle$, often referred to
as clock transition since its equivalent in cesium is used as the
primary time standard. Shortly after laser cooling had been
realized, the benefits of using ultracold atoms for atomic clocks
had become apparent \cite{kase89} and today the most accurate
atomic clocks are operated with laser-cooled atoms \cite{sant99}.
Therefore, it seems natural to investigate the use of a BEC with
its significantly reduced kinetic energy for the study of the
clock transition.

\begin{figure}
\epsfig{file=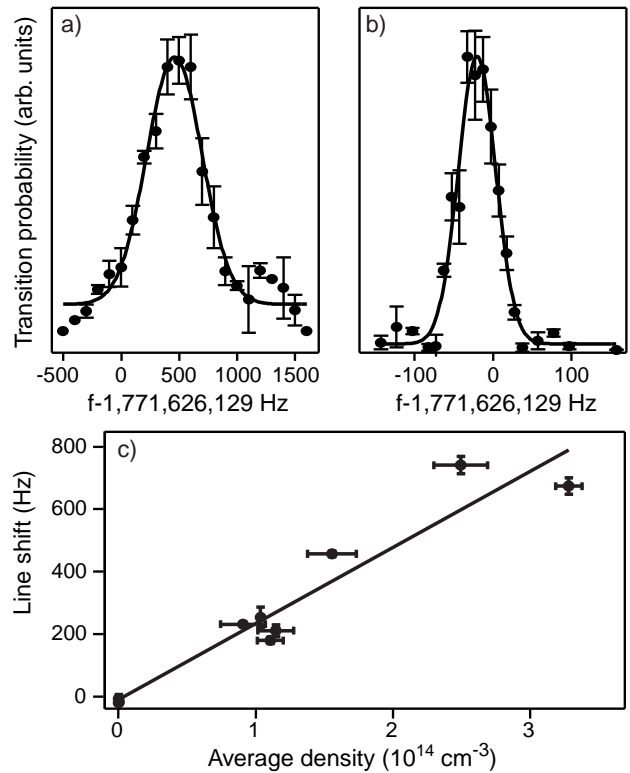,width = 3.25 in}
\caption{\label{fig:clock-transition}Magnetic-field insensitive
transition $|$1,0$\rangle$\,$\rightarrow$\,$|$2,0$\rangle$ in a
BEC. (a) Spectrum in the trap at a mean density of $1.6 \times
10^{14}$\,atoms/cm$^{3}$. (b) Spectrum after 12.5\,ms
time-of-flight at a mean density of $4.3 \times
10^{11}$\,atoms/cm$^{3}$. The discrepancy between the center of
the line and $\nu =0$ is probably due to an error in the exact
determination of the residual magnetic field. The solid lines are
Gaussian fits. (c) Transition frequency as a function of density
yielding a clock shift of $(2.44 \pm 0.25)\times
10^{-12}$\,Hz\,cm$^{3}$. }
\end{figure}

To observe the clock transition, we first completely transferred
an optically trapped $|$1,-1$\rangle$ condensate into the
$|$1,0$\rangle$ state with a radiofrequency Landau-Zener sweep.
Selective driving of the $|$1,-1$\rangle \, \rightarrow
\,|$1,0$\rangle$ transition was achieved  by applying a 3\,G
offset field which provided a large enough quadratic Zeeman-shift
to lift the degeneracy with the $|$1,0$\rangle \, \rightarrow
\,|$1,+1$\rangle$ transition. Subsequently, the magnetic field was
reduced to a value of typically 100\,mG which keeps the spins
aligned and gives rise to a quadratic Zeeman shift of the clock
transition of $\approx$\,20\,Hz. The $|$1,0$\rangle
\,\rightarrow\,|$2,0$\rangle$ transition was then excited by using
a microwave pulse at 1.77\,GHz with a duration between 2 and
5\,ms. The fraction of atoms transferred into the $|$2,0$\rangle$
state was kept below 20\,$\%$ in order to ensure a practically
constant density in the $|$1,0$\rangle$ state during the pulse.
Immediately afterwards, the optical trap was turned off suddenly
and the number of atoms which made the transition was detected by
state-selective absorption imaging after 15\,-\,30\,ms of
ballistic expansion. A typical spectrum showing the number of
transferred atoms as a function of microwave frequency (corrected
for the calculated quadratic Zeeman shift) for a BEC with an
average density of $1.6 \times 10^{14}$ atoms/cm$^3$ is shown in
Fig.\ \ref{fig:clock-transition}\,a). The density was determined
by measuring the release energy \cite{gorl011D2D} of $|$1,-1$
\rangle$ condensates without applying a microwave pulse. The
release energy $E_{rel}$ is related to the chemical potential
$\mu$ by $E_{rel}=(2/7) \mu=(2/7)(h^2 a_{|1,-1\rangle
|1,-1\rangle}/\pi m) n_o$ \cite{dalf99rmp}. Here, $a_{|a\rangle
|b\rangle}$ is the scattering length between two $^{23}$Na atoms
in states $|a\rangle$ and $|b\rangle$ ($a_{|1,-1\rangle
|1,-1\rangle}= 2.80$\,nm), $m$ is the $^{23}$Na mass, $h$ is
Planck's constant and $n_0$ is the peak density in the condensate
related to the average density by $\bar{n} = (4/7)\,n_0$. The
spectrum in Fig.\ \ref{fig:clock-transition}\,a) is significantly
broadened compared to the one in Fig.\
\ref{fig:clock-transition}\,b), which is taken after ballistic
expansion, and the transition frequency is shifted with respect to
the unperturbed frequency $\nu_0 = 1,771,626,129$\,Hz
\cite{kase89}.

In the limit of weak excitation, the density-dependent shift of
the clock-transition frequency is due to the difference in
mean-field potential that atoms in the $|$1,0$\rangle$ and
$|$2,0$\rangle$ state experience within a $|$1,0$\rangle$
condensate. Taking into account the inhomogeneous density
distribution of a trapped BEC, this leads to a line shape given by
\cite{sten99brag}

\begin{equation}
I(\nu) = \frac{15 h (\nu - \nu_0)}{4 n_0 \Delta U}
\sqrt{1- \frac{h (\nu - \nu_0)}{n_0 \Delta U} } \\
\label{equ:lineshape}
\end{equation}
\noindent with
\begin{equation}
 \Delta U  = \frac {h^2}{\pi m} (a_{|2,0\rangle
|1,0\rangle} - a_{|1,0\rangle |1,0\rangle}),
\label{equ:rel_potential}
\end{equation}

\noindent where the center of the line is at $\nu_0 + 2 n_0 \Delta
U/ 3 h$ and the average frequency is $\nu_0 +  4 n_0 \Delta U/ 7
h$. In our experiment, the line is additionally broadened and the
asymmetry of Eq.\ \ref{equ:lineshape} smeared out due to the
finite width of the microwave pulse which was limited by rapid
inelastic losses in the $|$2,0$\rangle$ state. Therefore, we have
used a (symmetric) Gaussian to fit the resonances where we have
identified the fitted center frequency as the average frequency of
the line. By taking spectra of the clock-transition at different
densities we have determined a density shift of $(2.44 \pm
0.25)\times 10^{-12}$\,Hz\,cm$^{3}$ (Fig.\
\ref{fig:clock-transition}\,c). Here, the error is the statistical
error from a linear fit to the data. Additional systematic errors
due to fitting of the line with a Gaussian and due to an
uncertainty in the determination of the density are estimated to
be smaller than 20\,$\%$. Using Eq.\ \ref{equ:rel_potential} and
$a_{|1,0\rangle |1,0\rangle} = 2.71$\,nm \cite{samu00}, we
determine the scattering length $a_{|2,0\rangle |1,0\rangle} =
3.15 \pm 0.05$\,nm for collisions between two atoms in states
$|$1,0$\rangle$ and $|$2,0$\rangle$.

In conclusion, we have prepared condensates in the upper F=2
hyperfine manifold  of the sodium ground state in a large-volume
optical trap and observed a stable condensate in the high-field
seeking stretched state $|$2,$-$2$\rangle$. Since only the
stretched state exhibits reasonable stability, experiments with
more complex spinor condensates do not seem to be possible.
Furthermore, we have for the first time observed the alkali
clock-transition in a Bose-Einstein condensate and determined the
value for the density-dependent mean-field shift. In present BEC
experiments, the magnitude of the shift precludes the use of
trapped condensates for precise atomic clocks. However, under
circumstances where the condensate density can be drastically
reduced as may be feasible in space-based experiments, the
extremely low velocity spread of BECs might help improve the
accuracy of atomic clocks.

This work was supported by NSF, ONR, ARO, NASA, and the David and
Lucile Packard Foundation. A.E.L. acknowledges additional support
from the NSF.


\end{document}